\documentclass[showpacs]{revtex4} %twocolumn
\usepackage{epsfig,amsmath,amssymb,graphicx,upgreek,textcomp}
\usepackage{natbib}
\usepackage[english]{babel} %

\begin{document}

\vspace{0mm}
\title{Thermodynamics of the Fermi gas in a cubic cavity of an arbitrary volume} %
\author{Yu.M. Poluektov}
\email{yuripoluektov@kipt.kharkov.ua (y.poluekt52@gmail.com)} %
\affiliation{National Science Center ``Kharkov Institute of Physics and Technology'', 61108 Kharkov, Ukraine} %
\author{A.A. Soroka} %
\affiliation{National Science Center ``Kharkov Institute of Physics and Technology'', 61108 Kharkov, Ukraine} %

\begin{abstract}
For the Fermi gas filling the space inside a cubic cavity of a fixed
volume, at arbitrary temperatures and number of particles, the
thermodynamic characteristics are calculated, namely: entropy,
thermodynamic potential, energy, pressure, heat capacities and
thermodynamic coefficients. The discrete structure of energy levels
is taken into account and size effects at low temperatures are
studied. The transition to the continual limit is considered.
\newline%
\end{abstract}

\maketitle

\vspace{10mm}
\section{Introduction}\vspace{-0mm} %\cite{}
The ideal Fermi gas model is the basis for understanding the
properties of metals, electron and other multifermion systems. In
many cases, it is possible to describe with acceptable accuracy also
the behavior of systems of interacting Fermi particles in the
framework of the approximation of an ideal gas of quasiparticles,
whose dispersion law differs from the dispersion law of free particles. %
It is essential that all thermodynamic characteristics of ideal
Fermi gas at arbitrary temperatures in the case of a large volume
can be expressed through the special Fermi-Stoner functions and,
thus, all relations of phenomenological thermodynamics can be
obtained and checked in the framework of the quantum microscopic model. %

Currently, much attention is paid to the study of quantum properties
of systems with a small number of particles, such as quantum dots,
other mesoscopic objects and nanostructures. In this connection, the
problem of description of such objects with taking into account
their interaction with the external environment is actual.

Statistical description is usually used to study systems with a very
large number of particles. But statistical methods of description
can also be applied in the study of equilibrium states of systems
with a small number of particles and even a single particle. When
considering a system within a large canonical ensemble, it is
assumed that it is a part of a very large system, a thermostat, with
which it can exchange energy and particles. The thermostat itself is
characterized by such statistical quantities as temperature $T$ and
chemical potential $\mu$. Assuming that the subsystem under
consideration is in thermodynamic equilibrium with the thermostat,
the subsystem itself, even consisting of a small number of
particles, will be characterized by the same quantities. For
example, we can consider the thermodynamics of an individual quantum
oscillator \cite{LL}. In the case when an exchange of particles with
a thermostat is possible, the time-averaged number of particles of a
small subsystem may be not an integer and, in particular, even less
than unity.

In statistical physics, entropy and distribution functions of
particles over quantum states are calculated under the assumption
that the number of particles is very large. Such consideration for
fermions leads to the Fermi-Dirac distribution, and for bosons -- to
the Bose-Einstein distribution \cite{LL}. In work \cite{PS}, the
authors calculated the entropy and distribution functions of
non-interacting particles in the case when no restrictions are
imposed on their number in a system being in thermodynamic
equilibrium with the environment. In \cite{PS2}, the thermodynamic
properties of a two-level system of finite volume were studied in detail. %

The influence of boundaries on the behavior of heat capacity in
metal colloids has been theoretically studied by Fr\"{o}hlich long
ago \cite{Fr}. Currently, the experimental possibilities for
studying low-dimensional systems at low temperatures are much wider,
so that there is an urgent need for a systematic, more detailed
theoretical study of such systems. Thermodynamic functions for the
Fermi gas confined between two planes and in a cylindrical tube were
previously calculated by the authors in \cite{PS3,PS4}.

In this work, for the Fermi gas filling the space inside a cubic
cavity of a fixed volume, at arbitrary temperatures and number of
particles, taking into account the results of \cite{PS,PS2}, we
calculate its thermodynamic characteristics, namely: entropy,
thermodynamic potential, energy, pressure, heat capacities and
thermodynamic coefficients. At low temperatures, the discrete
structure of energy levels is taken into account and size effects
are studied. The transition to the continual limit in the limit of
large volume and high temperatures is considered.

\section{States of Fermi particles in a cubic cavity}\vspace{-0mm} %
Let us consider the thermodynamic properties of the Fermi gas
enclosed in a cubic cavity with side $L$ and volume $V=L^3$. No
restrictions are imposed on the side length of the cube, and,
therefore, the approach used is applicable to the study of
small-sized objects containing a small number of particles.
Preliminarily we consider the classification of states of one
particle, assuming that its spin is equal to 1/2 and the potential
barrier on the surface of the cube is infinite, so that the wave
function of a particle at the boundaries turns to zero. Then its
normalized wave function has the form
\begin{equation} \label{01}
\begin{array}{l}
\displaystyle{%
  \psi(x,y,z)=\left(\frac{2}{L}\right)^{\!\!3\!/2} %
  \sin\!\Big(2\pi n_x\frac{x}{L}\Big)\sin\!\Big(2\pi n_y\frac{y}{L}\Big)\sin\!\Big(2\pi n_z\frac{z}{L}\Big). %
}%
\end{array}
\end{equation}
Wave numbers are determined by formula
$k_\alpha=\frac{2\pi}{L}n_\alpha$, and integer numbers run through
the values $n_\alpha=\pm 1, \pm 2\ldots$ Thus, each state is
characterized by a set of integers $(n_x,n_y,n_z)$, which for
brevity we will sometimes denote by a single symbol $\nu\equiv(n_x,n_y,n_z)$. %
The energy in state $\nu$
\begin{equation} \label{02}
\begin{array}{l}
\displaystyle{%
  \varepsilon_\nu=\frac{\hbar^2}{2m}\left(\frac{2\pi}{L}\right)^{\!\!2} %
  \big(n_x^2+n_y^2+n_z^2\big)=\frac{\hbar^2}{2m}\left(\frac{2\pi}{L}\right)^{\!\!2}\gamma^2 %
}%
\end{array}
\end{equation}
depends only on the combination of integers
\begin{equation} \label{03}
\begin{array}{l}
\displaystyle{%
  \gamma\equiv \sqrt{n_x^2+n_y^2+n_z^2}\,, %
}%
\end{array}
\end{equation}
so that the states $\nu$ are degenerate. Let us number the energy by
index $i=1,2,\ldots$ in the ascending order %
$\varepsilon_i=\frac{\hbar^2}{2m}\left(\frac{2\pi}{L}\right)^{2}\gamma_i^2$. %
Also denote by $z_i$ the multiplicity of degeneracy of a level,
taking into account the two-fold degeneracy in the spin projection
in the absence of a magnetic field. Thus, the summation over states
$\nu$ is equivalent to the summation over index $i$ with taking into
account the multiplicity of degeneracy: %
$\sum_\nu\,\ldots=\sum_{i=1}^\infty z_i\ldots$\,\,. %
The bottom ten states by the energy and the multiplicity of their
degeneracy are given in Table I.
\vspace{-4mm}%
\begin{table}[h!] \nonumber
\centering %
\caption{The bottom ten energy states in a cubic cavity} %
\vspace{0.5mm}%
\begin{tabular}{|c|c|c|c|c|c|c|c|c|c|c|} \hline  % \rule{3mm}{0pt} 0.385 \rule{3mm}{0pt}
\rule{3mm}{0pt} $i$          \rule{3mm}{0pt} & \rule{3mm}{0pt} 1 \rule{3mm}{0pt}  & \rule{3mm}{0pt} 2 \rule{3mm}{0pt}  & \rule{3mm}{0pt} 3 \rule{3mm}{0pt}  & \rule{3mm}{0pt} 4 \rule{3mm}{0pt}  & \rule{3mm}{0pt} 5 \rule{3mm}{0pt}  & \rule{3mm}{0pt} 6 \rule{3mm}{0pt} & \rule{3mm}{0pt} 7  \rule{3mm}{0pt} & \rule{3mm}{0pt} 8  \rule{3mm}{0pt}& \rule{3mm}{0pt} 9  \rule{3mm}{0pt}& \rule{3mm}{0pt} 10 \rule{3mm}{0pt} \\ \hline %
\rule{3mm}{0pt} $\gamma_i^2$ \rule{3mm}{0pt} & \rule{3mm}{0pt} 3 \rule{3mm}{0pt}  & \rule{3mm}{0pt} 6 \rule{3mm}{0pt}  & \rule{3mm}{0pt} 9 \rule{3mm}{0pt}  & \rule{3mm}{0pt} 11 \rule{3mm}{0pt} & \rule{3mm}{0pt} 12 \rule{3mm}{0pt} & \rule{3mm}{0pt} 14 \rule{3mm}{0pt}& \rule{3mm}{0pt} 17 \rule{3mm}{0pt} & \rule{3mm}{0pt} 18 \rule{3mm}{0pt}& \rule{3mm}{0pt} 19 \rule{3mm}{0pt}& \rule{3mm}{0pt} 21 \rule{3mm}{0pt} \\ \hline %
\rule{3mm}{0pt} $z_i$        \rule{3mm}{0pt} & \rule{3mm}{0pt} 16 \rule{3mm}{0pt} & \rule{3mm}{0pt} 48 \rule{3mm}{0pt} & \rule{3mm}{0pt} 48 \rule{3mm}{0pt} & \rule{3mm}{0pt} 48 \rule{3mm}{0pt} & \rule{3mm}{0pt} 16 \rule{3mm}{0pt} & \rule{3mm}{0pt} 96 \rule{3mm}{0pt}& \rule{3mm}{0pt} 48 \rule{3mm}{0pt} & \rule{3mm}{0pt} 48 \rule{3mm}{0pt}& \rule{3mm}{0pt} 48 \rule{3mm}{0pt}& \rule{3mm}{0pt} 96 \rule{3mm}{0pt} \\ \hline %
\end{tabular}  %tabular
\vspace{-3mm}
\end{table}
\newline\mbox{\hspace{18mm}} $i$ -- the level number in the order of increasing energy; $\gamma_i^2\equiv n_x^2+n_y^2+n_z^2$, %
\newline\mbox{\hspace{18mm}} $z_i$ -- the degeneracy factor of a level with account of two-fold degeneracy in the spin projection. %
\vspace{0mm}%

\section{Distribution function for arbitrary number of particles} \vspace{-1mm}%
Equations for the populations of levels for an arbitrary and even
small number of particles were obtained by the authors in
\cite{PS,PS2}. We reproduce here briefly these results for fermions.
When constructing the thermodynamics of a system with an arbitrary
number of particles and an arbitrary size, we will proceed from a
combinatorial expression for entropy. If at each level of a quantum
Fermi system with the energy $\varepsilon_j$ and degeneracy factor
$z_j$ there are $N_j$ particles, then the statistical weight of such
a state is given by the formula \cite{LL}
\begin{equation} \label{04}
\begin{array}{l}
\displaystyle{%
   \Delta\Gamma_j=\frac{z_j!}{N_j!\big(z_j-N_j\big)!}\,.  %
}
\end{array}
\end{equation}
The entropy is defined as the logarithm of the total statistical
weight by the relation
\begin{equation} \label{05}
\begin{array}{l}
\displaystyle{%
   S=\ln\Delta\Gamma =\sum_j\ln\Delta\Gamma_j= %
   \sum_j\!\Big[\ln z_j!-\ln N_j!-\ln\!\big(z_j-N_j\big)!\Big]. %
}
\end{array}
\end{equation}
To calculate all factorials for $N\gg 1$, the Stirling formula is
usually used in the form
\begin{equation} \label{06}
\begin{array}{l}
\displaystyle{%
    \ln N! \approx N \ln\!\Big(\frac{N}{e}\Big).  %
}
\end{array}
\end{equation}
For small $N$ the accuracy of this formula is insufficient. So, for
example, with $N=16$ its accuracy is 7.5\%. For $N=1,2$\, there are
negative numbers on the right in (6). If, as it is assumed, the
time-averaged number of particles can be arbitrary, in particular
small and fractional, the factorial should be defined through the
gamma function \cite{AS}:
\begin{equation} \label{07}
\begin{array}{l}
\displaystyle{%
   N! = \Gamma(N+1). %
}
\end{array}
\end{equation}
In this case, the statistical weight (4) is also expressed through
the gamma function:
\begin{equation} \label{08}
\begin{array}{l}
\displaystyle{%
   \Delta\Gamma_j = \frac{\Gamma(z_j+1)}{\Gamma(N_j+1)\Gamma(z_j-N_j+1)}. %
}
\end{array}
\end{equation}
This implies the formula for nonequilibrium entropy $S=\sum_jS_j$:
\begin{equation} \label{09}
\begin{array}{l}
\displaystyle{%
   S_j = \ln\Gamma(z_j+1) -\ln\Gamma(z_jn_j+1)-\ln\Gamma\big[z_j(1-n_j)+1\big]. %
}
\end{array}
\end{equation}
If a level is filled $n_j=1$ or empty $n_j=0$, then it does not
contribute to the total entropy. Taking into account that the total
number of particles $N$ and the total energy $E$ are determined by
the formulas
\begin{equation} \label{10}
\begin{array}{l}
\displaystyle{%
   N=\sum_jN_j=\sum_jn_jz_j,  %
}
\end{array}
\end{equation} \vspace{-4mm}
\begin{equation} \label{11}
\begin{array}{l}
\displaystyle{%
   E=\sum_j\varepsilon_jN_j=\sum_j\varepsilon_jn_jz_j,  %
}
\end{array}
\end{equation}
the average number of particles $n_j=N_j\big/z_j$ at level $j$, or
the population of the level, is found from the condition
\begin{equation} \label{12}
\begin{array}{l}
\displaystyle{%
   \frac{\partial}{\partial n_j}\big(S-\alpha N-\beta E\big)=0,  %
}
\end{array}
\end{equation}
where $\alpha,\beta$ are the Lagrange multipliers. From this
condition we find the equation that determines the average number of
particles in each state \cite{PS,PS2} at $0\le n_j\le 1$: %
\begin{equation} \label{13}
\begin{array}{l}
\displaystyle{%
   \theta_j\equiv\theta(n_j,z_j)\equiv\psi\big[z_j(1-n_j)+1\big] -\psi\big(z_jn_j+1\big) =\alpha+\beta\varepsilon_j=\frac{(\varepsilon_j-\mu)}{T},  %
}
\end{array}
\end{equation}
where $\psi(z)$  is the logarithmic derivative of the gamma function
(the psi function) \cite{AS}. From comparison with thermodynamic
relations it follows that $\alpha=-\mu/T$, $\beta=1/T$, $T$ --
temperature, $\mu$ -- chemical potential.  Using the asymptotic
formula $\psi(x)\sim\ln x -1/2x$, which is valid for $x\rightarrow\infty$, we obtain %
\begin{equation} \label{14}
\begin{array}{l}
\displaystyle{%
   n_j=n_j^{(0)}-\frac{\big(1-2n_j^{(0)}\big)}{2z_j}, %
}
\end{array}
\end{equation}
where
\begin{equation} \label{15}
\begin{array}{l}
\displaystyle{%
   n_j^{(0)}=\frac{1}{ \vphantom{^{1^1}} e^{(\varepsilon_j-\mu)\!/T}+1 } %
}
\end{array}
\end{equation}
is the usual Fermi-Dirac distribution. Formula (14) gives a good
approximation if $n_j$ is not close to zero or unity, and turns into
the usual distribution (15) in the limit $z_j\rightarrow\infty$. It
is important to note, however, that the exact function defined by
equation (13), in contrast to (15), becomes zero or unity at finite
values of energy.

\section{Thermodynamic functions, heat capacity, thermodynamic coefficients} %
The equilibrium entropy, number of particles and energy are
determined by formulas (9),\,(10),\,(11), taking into account that
the average number of particles at a given level as a function of
temperature and chemical potential is determined by equation (13).
The differential of the thermodynamic potential $\Omega=E-TS-\mu N$
has the usual form
\begin{equation} \label{16}
\begin{array}{l}
\displaystyle{%
   d\Omega=-SdT-Nd\mu-p\,dV. %
}
\end{array}
\end{equation}
The pressure
\begin{equation} \label{17}
\begin{array}{l}
\displaystyle{%
   p=-\sum_jz_jn_j\frac{d\varepsilon_j}{dV} %
}%
\end{array}
\end{equation}
is determined by the dependence of a particle's energy on volume. In
the case under consideration
$d\varepsilon_i\big/dV=-2\varepsilon_i/3V$, so that $p=2E/3V$. To
calculate heat capacities and thermodynamic coefficients, we first
find the differentials of the distribution function, number of
particles, entropy and pressure:
\begin{equation} \label{18}
\begin{array}{l}
\displaystyle{%
   z_jdn_j=\frac{\theta_j}{\theta_j^{(1)}}\frac{dT}{T}- %
           \frac{1}{\theta_j^{(1)}}\frac{d\varepsilon_j}{dV}\frac{dV}{T}+ %
           \frac{1}{\theta_j^{(1)}}\frac{d\mu}{T}, %
}%
\end{array}
\end{equation} \vspace{-4mm}
\begin{equation} \label{19}
\begin{array}{l}
\displaystyle{%
   dN=\frac{dT}{T}\sum_j\frac{\theta_j}{\theta_j^{(1)}} - %
      \frac{dV}{T}\sum_j\frac{1}{\theta_j^{(1)}}\frac{d\varepsilon_j}{dV} + %
      \frac{d\mu}{T}\sum_j\frac{1}{\theta_j^{(1)}}, %
}%
\end{array}
\end{equation} \vspace{-4mm}
\begin{equation} \label{20}
\begin{array}{l}
\displaystyle{%
   dS=\frac{dT}{T}\sum_j\frac{\theta_j^2}{\theta_j^{(1)}} - %
      \frac{dV}{T}\sum_j\frac{\theta_j}{\theta_j^{(1)}}\frac{d\varepsilon_j}{dV} + %
      \frac{d\mu}{T}\sum_j\frac{\theta_j}{\theta_j^{(1)}}, %
}%
\end{array}
\end{equation} \vspace{-4mm}
\begin{equation} \label{21}
\begin{array}{l}
\displaystyle{%
  dp=-\frac{dT}{T}\sum_j\frac{\theta_j}{\theta_j^{(1)}}\frac{d\varepsilon_j}{dV} + %
      \frac{dV}{T}\sum_j\!\left[\frac{1}{\theta_j^{(1)}}\!\left(\frac{d\varepsilon_j}{dV}\right)^{\!2}-z_jn_jT\,\frac{d^2\varepsilon_j}{dV^2}\right] - %
      \frac{d\mu}{T}\sum_j\frac{1}{\theta_j^{(1)}}\frac{d\varepsilon_j}{dV}. %
}%
\end{array}
\end{equation}
Here
\begin{equation} \label{22}
\begin{array}{l}
\displaystyle{%
   \theta_j^{(1)}\equiv\theta^{(1)}(n_j,z_j)\equiv\psi^{(1)}\big[z_j(1-n_j)+1\big] + \psi^{(1)}\big(z_jn_j+1\big), %
}
\end{array}
\end{equation}
where $\psi^{(1)}(y)=d\psi(y)/dy= d^{\,2}\ln\Gamma(y)\big/dy^2$ %
is the trigamma function \cite{AS}.

In the following we will consider systems with a fixed average
number of particles, for which $dN=0$. This condition allows to
eliminate the differential of chemical potential and, as a result,
the entropy and pressure differentials will take the form
\begin{equation} \label{23}
\begin{array}{l}
\displaystyle{%
   dS=B_T\frac{dT}{T} + B_V\frac{dV}{T}, \qquad  dp=A_T\frac{dT}{T} + A_V\frac{dV}{T}, %
}
\end{array}
\end{equation}
where
\begin{equation} \label{24}
\begin{array}{l}
\displaystyle{%
  B_T=\sum_j\frac{\theta_j^2}{\theta_j^{(1)}}-\theta^{(1)}\bigg(\sum_j\frac{\theta_j}{\theta_j^{(1)}}\bigg)^{\!2}, \quad%
  B_V=A_T=\theta^{(1)}\bigg(\sum_j\frac{\theta_j}{\theta_j^{(1)}}\bigg)\!\bigg(\sum_j\frac{1}{\theta_j^{(1)}}\frac{d\varepsilon_j}{dV}\bigg) - %
           \bigg(\sum_j\frac{\theta_j}{\theta_j^{(1)}}\frac{d\varepsilon_j}{dV}\bigg), %
}\vspace{3mm}\\ %
\displaystyle{\hspace{0mm}%
  A_V=-T\sum_j z_jn_j\frac{d^2\varepsilon_j}{dV^2}-\theta^{(1)}\bigg(\sum_j\frac{1}{\theta_j^{(1)}}\frac{d\varepsilon_j}{dV}\bigg)^{\!2} + %
  \sum_j\frac{1}{\theta_j^{(1)}}\bigg(\frac{d\varepsilon_j}{dV}\bigg)^{\!2}, \quad %
  \frac{1}{\theta^{(1)}}\equiv \sum_j\frac{1}{\theta_j^{(1)}}\,.
}%
\end{array}
\end{equation}
In our case $d\varepsilon_i/dV=-2\varepsilon_i/3V$,
$d^2\varepsilon_i/dV^2=10\varepsilon_i/9V^2$. From these relations
there follow the formulas for the isochoric
\begin{equation} \label{25}
\begin{array}{l}
\displaystyle{%
   C_V=T\bigg(\frac{\partial S}{\partial T}\bigg)_{\!N,V}=B_T,  %
}%
\end{array}
\end{equation}
and the isobaric
\begin{equation} \label{26}
\begin{array}{l}
\displaystyle{%
   C_p=T\bigg(\frac{\partial S}{\partial T}\bigg)_{\!N,p}=B_T-\frac{B_V^2}{A_V} %
}%
\end{array}
\end{equation}
heat capacities. We also present following from (23),\,(24) formulas
for the coefficient of volumetric expansion
\begin{equation} \label{27}
\begin{array}{l}
\displaystyle{%
   \alpha_p=\frac{1}{V}\bigg(\frac{\partial V}{\partial T}\bigg)_{\!p}=-\frac{B_V}{VA_V},  %
}%
\end{array}
\end{equation}
the isothermal compressibility
\begin{equation} \label{28}
\begin{array}{l}
\displaystyle{%
  \gamma_T=-\frac{1}{V}\bigg(\frac{\partial V}{\partial p}\bigg)_{\!T}=-\frac{T}{VA_V}  %
}%
\end{array}
\end{equation}
and the isochoric thermal pressure coefficient
\begin{equation} \label{29}
\begin{array}{l}
\displaystyle{%
   \beta_V=\frac{1}{p}\bigg(\frac{\partial p}{\partial T}\bigg)_{\!V}=\frac{B_V}{p\,T}.  %
}%
\end{array}
\end{equation}
All other thermodynamic coefficients can be expressed through the
heat capacities and the coefficients given here \cite{RR}.
Obviously, the general relation \cite{LL}
\begin{equation} \label{30}
\begin{array}{l}
\displaystyle{%
   C_p-C_V=TV\frac{\alpha_p^2}{\gamma_T}=-\frac{B_V^2}{A_V}  %
}%
\end{array}
\end{equation}
is fulfilled, which confirms the consistency of the given general
thermodynamic relations. Since the thermodynamic inequalities
$C_V>0$ and $\big(\partial p/\partial V\big)_T<0$ must be satisfied
in a stable system \cite{LL}, then there must be $B_T>0$ and $A_V<0$. %

\section{Low temperatures. Size effects} %
The most interesting is the case when there are a small number of
particles in the volume of a cube of small size. In this case, the
discrete structure of the spectrum turns out to be important, so
that the exact formulas (9)\,--\,(11),\,(13),\,(17),\,(25)\,--\,(29)
should be used for calculations. Thermodynamic properties of such a
system of particles with two discrete levels are considered in
detail by the authors in \cite{PS2}. In the following, along with
dimensional quantities, we will use the writing of quantities in
dimensionless form, introducing arbitrary characteristic scales of
length $a_*$, energy
$\varepsilon_*=\big(\hbar^2/2m\big)\big(2\pi/a_*\big)^2$ and
pressure $p_*\equiv \big(2\pi^{3/2}\varepsilon_*/a_*^3\big)$. Note
that $a_*^2\varepsilon_*=\big(L^2\varepsilon_1/\gamma_1^2\big)$. For
example, the Bohr radius $a_*=a_B=0.53\cdot 10^{-8}$\,cm or some
other length can be chosen as the spatial scale. The dimensionless
length $\tilde{L}$, temperature $\tau$, pressure $\tilde{p}$ and
level energy $\tilde{\varepsilon}_j$ are defined by the relations:
\begin{equation} \label{31}
\begin{array}{l}
\displaystyle{%
   \tilde{L}=\frac{L}{a_*}, \quad \tau=\frac{T}{\varepsilon_*}, \quad %
   \tilde{p}=\frac{p}{p_*}, \quad \tilde{\varepsilon}_j=\frac{\varepsilon_j}{\varepsilon_*}=\frac{\gamma_j^2}{\tilde{L}^2}. %
}%
\end{array}
\end{equation}

Let us estimate the temperature and the cube size, at which the size
effects become significant. Obviously, these effects will manifest
themselves when the temperature becomes comparable to the energy
difference between levels, which is of the order of the energy of
the ground level. Its energy can be represented in the form
\begin{equation} \label{32}
\begin{array}{l}
\displaystyle{%
   \varepsilon_1=\frac{2\pi^2\hbar^2}{mL^2}\,\gamma_1^2=12\pi^2\bigg(\frac{a_B}{L}\bigg)^{\!2}\rm{Ry},  %
}%
\end{array}
\end{equation}
where $a_B=0.53\cdot 10^{-8}$\,cm, $\rm{Ry}=13.6$\,eV\,$=1.6\cdot
10^{5}$\,K is the Rydberg unit. The size, at which
$\varepsilon_1\approx 1$\,K is equal to $a_B/L\approx 0.2\cdot
10^{-3}$ or $L\approx 2.5\cdot 10^{-5}$\,cm. Thus, at temperature of
the order of one Kelvin the size effects should manifest themselves
already in samples of macroscopic dimensions.

Let us first consider the state of the system at zero temperature.
If the number of particles is less than or equal to the degeneracy
factor of the first level $0<N\le z_1$, then $\mu=\varepsilon_1$,
$0<n_1\le 1$ and the particles are only at the ground level. Higher
levels are not filled. In this case the population of the first
level $n_1$ is determined by the relation $N=z_1n_1$, and the
energy, pressure and thermodynamic coefficients are equal to
$E=\varepsilon_1z_1n_1$, $p=2E/3V$, $\alpha_p=0$, $\beta_V=0$,
$\gamma_T=(3/5)p^{-1}$. The entropy
\begin{equation} \label{33}
\begin{array}{l}
\displaystyle{%
   S = \ln\Gamma(z_1+1) -\ln\Gamma(z_1n_1+1)-\ln\Gamma\big[z_1(1-n_1)+1\big] %
}
\end{array}
\end{equation}
turns to zero only for the fully occupied level $n_1=1$ and is
different from zero for the unfilled level. In this case, the third
law of thermodynamics is satisfied in the Nernst formulation, while
in the Planck formulation it is satisfied only at fully occupied levels. %

If $M-1$ lower levels are completely filled, and the level $M$ can
be partially filled, then
$\sum_{j=1}^{M-1}\!z_j<N\le\sum_{j=1}^{M-1}\!z_j\,+z_M$ and
$0<n_M\le 1$, and the chemical potential $\mu=\varepsilon_M$. In
this case, the entropy is given by formula (33) with the
substitution $n_1\rightarrow n_M$, and the total number of
particles, energy, pressure and thermodynamic coefficients are
determined by formulas %
$N=\sum_{j=1}^{M-1}\!z_j \,+n_Mz_M$,
$E=\sum_{j=1}^{M-1}\!\varepsilon_jz_j \,+\varepsilon_Mn_Mz_M$,
$p=2E/3V$, $\alpha_p=0$, $\beta_V=0$, $\gamma_T=(3/5)p^{-1}$. %
The size effect, which manifest itself in discreteness of levels,
leads to the situation that near zero temperature there is a
temperature region in which the populations of levels do not change
with temperature and remain the same as at $T=0$.

In the temperature region where the temperature dependence of
populations arises, the quantities (24) through which the heat
capacities and thermodynamic coefficients (25)\,--\,(30) are
expressed, can be represented in dimensionless variables in the form
\begin{equation} \label{34}
\begin{array}{l}
\displaystyle{%
   B_T=\frac{G}{\tau^2\tilde{L}^4}, \qquad B_V=A_T=\frac{2\varepsilon_*}{3\tau\tilde{L}^7a_*^3}\,G, \qquad %
   A_V=-\frac{4\varepsilon_*^2}{9\tilde{L}^8a_*^6}\bigg(\frac{5}{2}\,\tau d - \frac{G}{\tilde{L}^2} \bigg),  %
}%
\end{array}
\end{equation}
where the notation is used
\begin{equation} \label{35}
\begin{array}{cc}
\displaystyle{%
  G\equiv g_4 - \theta^{(1)}g_2^2, %
}\vspace{3mm}\\ %
\displaystyle{\hspace{0mm}%
  g_2\equiv\sum_j\frac{\gamma_j^2}{\theta^{(1)}(n_j,z_j)}, \quad %
  g_4\equiv\sum_j\frac{\gamma_j^4}{\theta^{(1)}(n_j,z_j)}, \quad %
  \frac{1}{\theta^{(1)}}\equiv\sum_j\frac{1}{\theta^{(1)}(n_j,z_j)}, \quad %
  d\equiv\sum_j\gamma_j^2\,z_j\,n_j.
}%
\end{array}
\end{equation}
Note that the quantity $G$ does not explicitly contain the chemical
potential and is positive due to the requirement of thermodynamic
stability. The coefficient $A_V$ in a thermodynamically stable
system must be negative $A_V<0$. It entails the necessity of
fulfillment of the inequality $\tau>\tau_*$, where
\begin{equation} \label{36}
\begin{array}{l}
\displaystyle{%
   \tau_*\equiv\frac{2}{5}\frac{G}{\tilde{L}^2d}.
}%
\end{array}
\end{equation}
At $\tau<\tau_*$ the excited system is unstable, and so the system
continues to remain in the same ground state as at $T=0$. Thus, the
calculation of the temperature dependence of any thermodynamic
quantity is reduced to the calculation of sums (35). In particular,
the pressure (17) is
\begin{equation} \label{37}
\begin{array}{l}
\displaystyle{%
   \frac{p}{p_*}=\frac{d}{3\pi^{3/2}\tilde{L}^5}.
}%
\end{array}
\end{equation}

As noted, we consider systems in which the time-averaged number of
particles can vary within range $0<N<\infty$. As will be seen, there
is some critical value $N_*$, so that the cases $N>N_*$ and $N<N_*$
should be considered separately. Let us first consider the case
$N>N_*$, assuming that the number of particles is less than or equal
to the degeneracy factor of the first level $N\le z_1$. In this
case, particles will begin to transit from the ground to the second
level at the temperature
\begin{equation} \label{38}
\begin{array}{l}
\displaystyle{%
   \tau_0=\frac{\big(\gamma_2^2-\gamma_1^2\big)}{\tilde{L}^2Z_0}.
}%
\end{array}
\end{equation}
In case of fulfillment of the conditions $N>N_*$ and $\tau\ge
\tau_0$, the stability condition $A_V<0$ is satisfied. At the
critical value of $N=N_*\approx 0.178$ the coefficient
$A_V(\tau_0;N_*)=0$ turns to zero already at the temperature
$\tau_0$, and at $\tau>\tau_0$ it proves to be positive. Therefore,
the case $N<N_*$ will be considered separately below.

Here and further we use the notation
\begin{equation} \label{39}
\begin{array}{l}
\displaystyle{%
  Z_0\equiv \theta(0,z_2)-\theta(N/z_1,z_1), \quad %
  Z_1\equiv \theta^{(1)}(0,z_2)+\theta^{(1)}(N/z_1,z_1), \quad %
  Z_2\equiv \theta^{(2)}(0,z_2)-\theta^{(2)}(N/z_1,z_1). %
}%
\end{array}
\end{equation}
The functions $\theta(n_j,z_j), \theta^{(1)}(n_j,z_j)$  are defined
in (13),\,(22), and
\begin{equation} \label{40}
\begin{array}{l}
\displaystyle{%
   \theta_j^{(2)}(n_j,z_j)=\psi^{(2)}\big[z_j(1-n_j)+1\big] - \psi^{(2)}\big(z_jn_j+1\big), %
}
\end{array}
\end{equation}
where $\psi^{(2)}(x)=d^2\psi(x)\big/dx^2$. In the considered case
$N>N_*$ the inequality $\tau_0>\tau_*$ is fulfilled, so as noted
above the stability condition $A_V<0$ at $\tau_0$ is satisfied.
Obviously, the existence of characteristic temperatures (36),\,(38)
is caused by the finite size of the system and they tend to zero at
$\tilde{L}\rightarrow\infty$. At temperature slightly above (38)
$\tau=\tau_0+0$ we have
\begin{equation} \label{41}
\begin{array}{l}
\displaystyle{%
  G\equiv G_0=\frac{\big(\gamma_2^2-\gamma_1^2\big)^2}{Z_1}, \qquad %
  d\equiv d_0=\gamma_1^2N.
}%
\end{array}
\end{equation}
Taking this into account we conclude that the quantities $E,p,S$ are
continuous at $\tau_0$, and the heat capacities and thermodynamic
coefficients undergo jumps here:
\begin{equation} \label{42}
\begin{array}{cc}
\displaystyle{%
  \Delta C_V\equiv C_{V+}-C_{V-}=\frac{Z_0^2}{Z_1}, \qquad %
  \Delta C_p\equiv C_{p+}-C_{p-}=\frac{Z_0^2}{Z_1}\bigg[1+\big(\gamma_2^2-\gamma_1^2\big)\frac{Z_0}{Z_1}A_2^{-1}\bigg],  %
}\vspace{3mm}\\ %
\displaystyle{\hspace{0mm}%
  \Delta\alpha_p= \alpha_{p+}-\alpha_{p-}=\frac{3}{2}\frac{\gamma_1^2}{\varepsilon_1}\frac{Z_0^2}{Z_1A_2}, \qquad %
  \Delta\gamma_T= \gamma_{T+}-\gamma_{T-}=\frac{9}{10}\frac{L^3}{N\varepsilon_1}\bigg(\frac{5}{2}\frac{N\gamma_1^2}{A_2}-1\bigg),  %
}\vspace{3mm}\\ %
\displaystyle{\hspace{0mm}%
  \Delta\beta_V= \beta_{V+}-\beta_{V-}=\frac{1}{N\varepsilon_1}\frac{Z_0^2}{Z_1}, %
}%
\end{array}
\end{equation}
where %
$\displaystyle{A_2\equiv\frac{5}{2}\,\gamma_1^2N \!-\! \big(\gamma_2^2-\gamma_1^2\big)\frac{Z_0}{Z_1}}$,  %
and $C_{V+}\equiv C_V(\tau_0+0)$, $C_{V-}\equiv C_V(\tau_0-0)$ et al. %

Let us consider the behavior of thermodynamic quantities near and
above the temperature (38), assuming $\tau=\tau_0+\Delta\tau$, where
$\Delta\tau/\tau_0\ll 1$. Here the continuous transition of
particles from the ground to the second level begins. Then the
populations of levels are $n_1=N/z_1+\Delta n_1$, $n_2=\Delta n_2$,
and also $z_1\Delta n_1/N\ll 1$. From the condition of conservation
of the total number of particles it follows that $z_1\Delta
n_1+z_2\Delta n_2=0$. The dependence of populations on temperature
is found from the system of equations
\begin{equation} \label{43}
\begin{array}{l}
\displaystyle{%
  \theta(n_1,z_1)=\frac{\gamma_1^2}{\tau\tilde{L}^2}-t, \qquad %
  \theta(n_2,z_2)=\frac{\gamma_2^2}{\tau\tilde{L}^2}-t, \qquad %
}%
\end{array}
\end{equation}
where $t=\mu/T$. As a result, near and above $\tau_0$ we have
\begin{equation} \label{44}
\begin{array}{l}
\displaystyle{%
  z_1\Delta n_1=-z_2\Delta n_2=-\frac{Z_0}{Z_1}\frac{\Delta\tau}{\tau_0}, %
}%
\end{array}
\end{equation} \vspace{-4mm}
\begin{equation} \label{45}
\begin{array}{l}
\displaystyle{%
  G\approx G_0+\Delta G, \quad \Delta G=\frac{Z_0Z_2}{Z_1^2}\,G_0\,\frac{\Delta\tau}{\tau_0}, %
}%
\end{array}
\end{equation}\vspace{-4mm}
\begin{equation} \label{46}
\begin{array}{l}
\displaystyle{%
  d\approx d_0+\Delta d, \quad \Delta d= \big(\gamma_2^2-\gamma_1^2\big)\frac{Z_0}{Z_1}\frac{\Delta\tau}{\tau_0}. %
}%
\end{array}
\end{equation}
From (44)\,--\,(46) and relations (9),\,(25)\,--\,(29),\,(37) it
follows that near and above $\tau_0$ the thermodynamic quantities
$\Theta\equiv\big\{S,C_V,C_p,p,\alpha_p,\gamma_T,\beta_V\big\}$
depend linearly on temperature:
\begin{equation} \label{47}
\begin{array}{l}
\displaystyle{%
  \frac{\Theta-\Theta_0}{\Theta_0}=K\,\frac{\Delta\tau}{\tau_0}. %
}%
\end{array}
\end{equation}
The coefficient $K$, which determines the slope of the line, for
different quantities is given by the formulas:
\begin{equation} \label{48}
\begin{array}{ccc}
\displaystyle{%
  K_S=\frac{Z_0^2}{S_0Z_1}, \quad%
  K_p=\frac{\big(\gamma_2^2-\gamma_1^2\big)}{\gamma_1^2N}\frac{Z_0}{Z_1}, %
}\vspace{3mm}\\ %
\displaystyle{\hspace{0mm}%
  K_{C_V}=\frac{Z_0Z_2}{Z_1^2}-2, \quad %
  K_{C_p}=\frac{Z_0^2}{C_{p0}Z_1}\bigg\{\!\bigg(\frac{Z_0Z_2}{Z_1^2}-2\bigg) +   %
  \big(\gamma_2^2-\gamma_1^2\big)\frac{Z_0}{A_2Z_1} %
  \bigg[2\bigg(\frac{Z_0Z_2}{Z_1^2}-1\bigg)-\frac{A_1}{A_2}\bigg]\bigg\},
}\vspace{3mm}\\ %
\displaystyle{\hspace{0mm}%
  K_{\alpha_p}=\frac{Z_0Z_2}{Z_1^2}-1-\frac{A_1}{A_2}, \quad %
  K_{\gamma_T}=1-\frac{A_1}{A_2}, \quad %
  K_{\beta_V}=\frac{Z_0Z_2}{Z_1^2}-2-\frac{\big(\gamma_2^2-\gamma_1^2\big)}{\gamma_1^2N}\frac{Z_0}{Z_1}. %
}%
\end{array}
\end{equation}
Here %
$\displaystyle{A_1\equiv\frac{5}{2}\,\gamma_1^2N+\!\big(\gamma_2^2-\gamma_1^2\big) %
\frac{Z_0}{Z_1}\bigg[\frac{5}{2}-\frac{Z_0Z_2}{Z_1^2}\bigg]}$, %
and the values of thermodynamic quantities at temperature $\tau_0+0$
are determined by the formulas:
\begin{equation} \label{49}
\begin{array}{ccc}
\displaystyle{%
   S_0 = \ln\Gamma(z_1+1) -\ln\Gamma(N+1)-\ln\Gamma\big[z_1-N+1\big], %
}\vspace{3mm}\\ %
\displaystyle{\hspace{0mm}%
  C_{V0}=\frac{Z_0^2}{Z_1}, \quad %
  C_{p0}=\frac{Z_0^2}{Z_1}+\big(\gamma_2^2-\gamma_1^2\big)\frac{Z_0^3}{Z_1^2}A_2^{-1}, %
}\vspace{3mm}\\ %
\displaystyle{\hspace{0mm}%
  \alpha_{p0}=\frac{3}{2}\frac{\gamma_1^2}{\varepsilon_1}\frac{Z_0^2}{Z_1A_2}, \quad %
  \gamma_{T0}=\frac{9}{4}\frac{\gamma_1^2L^3}{\varepsilon_1A_2}, \quad %
  \beta_{V0}=\frac{1}{N\varepsilon_1}\frac{Z_0^2}{Z_1}. %
}%
\end{array}
\end{equation}
The coefficients (48) and their sign depend on the number of
particles. The coefficients $K_S$ and $K_p$ are positive for all
$N>N_*$. The dependences of the coefficients $K_{C_V}$ and $K_{C_p}$
on the number of particles are shown in Fig.\,1. The coefficient
$K_{C_V}$ at $N_*$ has a finite negative value
$K_{C_V}\big(N_*\big)=-1.87$, and at $N=3.674$ it changes sign. The
coefficient $K_{C_p}$ tends to an infinite negative value at
$N\rightarrow N_*+0$ and changes sign at $N=3.368$.
\begin{figure}[h!]
\vspace{-2mm}  \hspace{0mm}
\includegraphics[width = 7.35cm]{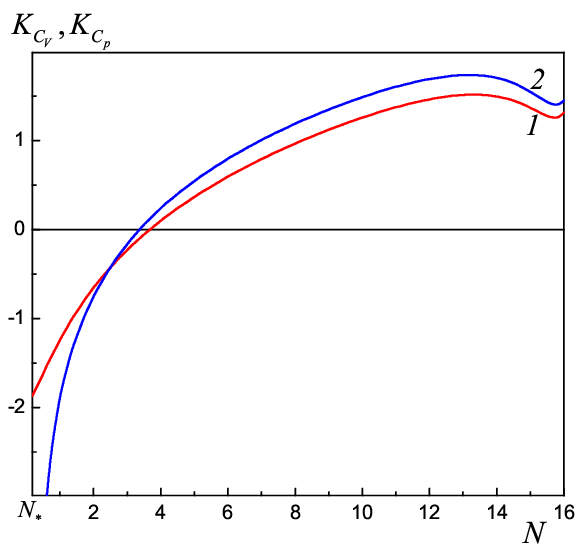} % 1.0\columnwidth
\vspace{-4mm} %
\caption{\label{fig01} %\hspace{10mm} %
Dependences of the slope coefficients $K_{C_V}(N)$ -- {\it 1} and
$K_{C_p}(N)$ -- {\it 2} on the number of particles at $N_*\le N\le z_1$. %
}%
\end{figure}
The dependences of the coefficients $K_{\alpha_p}$, $K_{\beta_V}$
and $K_{\gamma_T}$ on the number of particles are shown in Fig.\,2.
They are qualitatively similar to those shown in Fig.\,1. The
coefficient $K_{\beta_V}$ at $N_*$ takes on a finite negative value
$K_{\beta_V}\big(N_*\big)=-4.37$ and changes sign at $N=5.06$. The
coefficients $K_{\alpha_p}$ and $K_{\gamma_T}$ tend to an infinite
negative value at $N\rightarrow N_*+0$. The coefficient
$K_{\alpha_p}$ changes sign at $N=4.52$, and $K_{\gamma_T}$ turns to
zero at the points $N_1=12.59,\,N_2=13.95$. All curves in figures 1
and 2 have minima at $N$ near $z_1=16$.
\begin{figure}[h!]
\vspace{-2mm}  \hspace{0mm}
\includegraphics[width = 7.35cm]{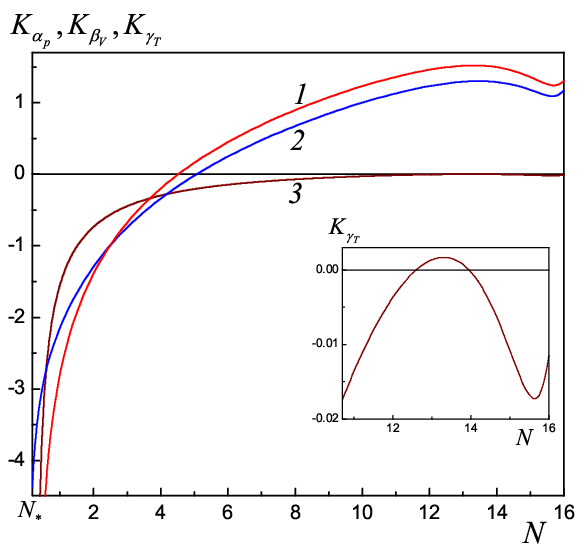} % 1.0\columnwidth
\vspace{-4mm} %
\caption{\label{fig02} %\hspace{10mm} %
Dependences of the slope coefficients $K_{\alpha_p}(N)$ -- {\it 1},
$K_{\beta_V}(N)$ -- {\it 2}, $K_{\gamma_T}(N)$ -- {\it 3} on the
number of particles at $N_*\le N\le z_1$. In the inset: the
dependence $K_{\gamma_T}(N)$ in the enlarged scale near the abscissa axis. %
}%
\end{figure}

With a further increase in temperature there occur transitions of
particles not only to the second, but also to higher levels, so that
the temperature dependences of the quantities become more complex.
As an illustration, Fig.\,3 shows the calculation of the temperature
dependences of populations of levels up to a temperature close to
the energy of the fourth level for $N=1$ under the assumption
$\tilde{n}=na_*^3=1$. At temperature $\tau_2\equiv\tau_0$ the
transition of particles from the first to the second level begins,
at $\tau_3,\tau_4$ the third and fourth levels begin to populate
with particles, respectively.

At $N<N_c$, where $N_c\approx 3.27$, there is a characteristic
temperature $\tau_{0,1}$ at which the probability of filling of the
first level turns to zero. At $N\rightarrow N_c-0$ the temperature
$\tau_{0,1}\rightarrow\infty$, with $n_2\approx n_3\approx n_4$. At
$N>N_c$ the temperature $\tau_{0,1}$ is absent, and some fraction of
particles continues to remain at the ground level at all
temperatures.

\begin{figure}[h!]
\vspace{0mm}  \hspace{0mm}
\includegraphics[width = 7.5cm]{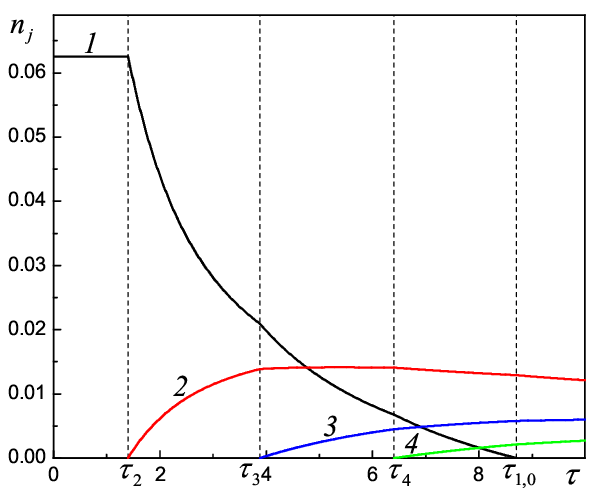} % 1.0\columnwidth
\vspace{-4mm} %
\caption{\label{fig03} %\hspace{10mm} %
The temperature dependences of populations of levels $n_j(\tau)$ at
$j=1,2,3,4$. The temperatures $\tau_2\equiv\tau_0=1.402$,
$\tau_3=3.885$, $\tau_4=6.403$ correspond to the beginning of
filling of levels 2,\,3 and 4, and at $\tau_{0,1}=8.700$ the
population of the ground level turns to zero. The energy of the
fourth level $\tilde{\varepsilon}_4=11.0$, $N=1$, $\tilde{n}=na_*^3=1$. %
}%
\vspace{-4mm}
\end{figure}

Figure\,4 shows the dependences of heat capacities $C_V(\tau)$ and
$C_p(\tau)$ on temperature under the same conditions as in Fig.\,3.
The heat capacities undergo jumps at temperatures at which the
filling of new levels begins, or when the population of the ground
level turns to zero.
\begin{figure}[h!]
\vspace{-2mm}  \hspace{0mm}
\includegraphics[width = 7.35cm]{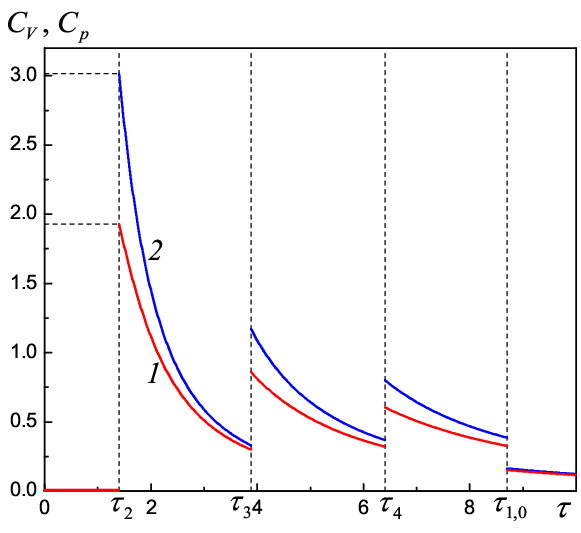} % 1.0\columnwidth
\vspace{-4mm} %
\caption{\label{fig04} %\hspace{10mm} %
The temperature dependencies of the heat capacities $C_V(\tau)$ --
{\it 1} and $C_p(\tau)$ -- {\it 2}. The heat capacities undergo
jumps at the temperatures
$\tau_2\equiv\tau_0,\,\tau_3,\,\tau_4,\,\tau_{1,0}$. A calculation
is carried out with the same parameters as in Fig.\,3.
}\vspace{-1.5mm}%
\end{figure}

The developed approach allows to consider systems in which the
time-averaged number of particles, due to the exchange of particles
with the thermostat, can be non-integer, and, in particular, less
than unity. As noted above, the case $N<N_*=0.178$ should be
considered separately, because in this case the system at $\tau_0$
is unstable due to the inequality $\tau_0<\tau_*$. In this case to
be considered now, the system remains in the ground state up to the
temperature $\tau_*$. When going above this temperature, a
redistribution of particles between two levels to the values
$n_{1*}$ and $n_{2*}$ occurs by a jump.

In what follows we will use notations similar to (39):
\begin{equation} \label{50}
\begin{array}{l}
\displaystyle{%
  \tilde{Z}_0\equiv \theta(n_{2*},z_2)-\theta(n_{1*},z_1), \quad %
  \tilde{Z}_1\equiv \theta^{(1)}(n_{2*},z_2)+\theta^{(1)}(n_{1*},z_1), \quad %
  \tilde{Z}_2\equiv \theta^{(2)}(n_{2*},z_2)-\theta^{(2)}(n_{1*},z_1), %
}%
\end{array}
\end{equation}
and also $\tilde{d}\equiv\gamma_1^2z_1n_1 + \gamma_2^2z_2n_2$. From
(36), taking into account that $\tilde{d}_*\equiv\gamma_1^2z_1n_{1*}
+ \gamma_2^2z_2n_{2*}$, and $G\equiv
G_*=\big(\gamma_2^2-\gamma_1^2\big)^2\big/\tilde{Z}_1$, it follows
that the temperature $\tau_*$ is determined by the equation
\begin{equation} \label{51}
\begin{array}{l}
\displaystyle{%
   \tau_*=\frac{2}{5}\frac{\big(\gamma_2^2-\gamma_1^2\big)^2}{\tilde{L}^2\tilde{Z}_1\tilde{d}_*}, %
}%
\end{array}
\end{equation}
and besides $\big(z_1n_{1*}/N\big)+\big(z_2n_{2*}/N\big)=1$. The
last relation is fulfilled if we introduce the angle $\alpha$, such that %
$\big(z_1n_{1*}/N\big)\equiv\cos^2\alpha$, $\big(z_2n_{2*}/N\big)\equiv\sin^2\alpha$. %
From equations (43) we find %
\begin{equation} \label{52}
\begin{array}{l}
\displaystyle{%
   \tau_*=\frac{\big(\gamma_2^2-\gamma_1^2\big)}{\tilde{L}^2\tilde{Z}_0}. %
}%
\end{array}
\end{equation}
By eliminating $\tau_*$ from (51) and (52), we obtain the equation
that allows to find the populations $n_{1*}$ and $n_{2*}$ at $\tau=\tau_*+0$: %
\begin{equation} \label{53}
\begin{array}{l}
\displaystyle{%
   \frac{2}{5}\frac{\big(\gamma_2^2-\gamma_1^2\big)\tilde{Z}_0}{\tilde{d}_*\,\tilde{Z}_1}=1. %
}%
\end{array}
\end{equation}
Knowing the populations, from (51) or (52) we find the critical
temperature $\tau_*$. The calculation for $N=0.1$ gives
$\tau_*=12.56$, $n_{1*}=0.0037$, $n_{2*}=0.00085$. Equation (53) has
a solution in the range of the number of particles $N_m\le N\le N_*$. %
Here $N_*=0.178$, to which corresponds $\alpha=0$, $n_{1*}=N_*/z_1$,
$n_{2*}=0$, and $N_m=0.061$, to which corresponds $\alpha=\pi/2$,
$n_{1*}=0$, $n_{2*}=N_m/z_2$. The dependencies of the temperatures
$\tau_0(N)$ and $\tau_*(N)$ on the number of particles are shown in Fig.\,5. %
\begin{figure}[h!]
\vspace{-2mm}  \hspace{0mm}
\includegraphics[width = 7.35cm]{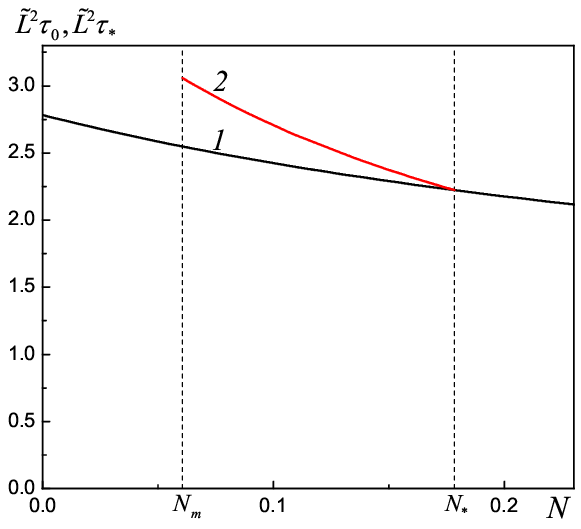} % 1.0\columnwidth
\vspace{-4mm} %
\caption{\label{fig05} %\hspace{10mm} %
Dependencies of the characteristic temperatures
$\tilde{L}^2\tau_0(N)$ -- {\it 1}, $\tilde{L}^2\tau_*(N)$ -- {\it 2}
on the number of particles. %
At $N_m=0.061$: $\tau_*\tilde{L}^2=3.057$, $\tau_0\tilde{L}^2=2.548$. %
At $N_*=0.178$: $\tau_*\tilde{L}^2=\tau_0\tilde{L}^2=2.223$.
}\vspace{-1.5mm}%
\end{figure}

At temperature $\tau_*$ the energy, pressure, entropy, isochoric
heat capacity and isochoric thermal pressure coefficient undergo a jump %
\begin{equation} \label{54}
\begin{array}{ccc}
\displaystyle{%
  \Delta E=E_+-E_-=\varepsilon_1\bigg(\frac{\gamma_2^2}{\gamma_1^2}-1\bigg)z_2n_{2*}, \quad %
  \Delta p=p_+-p_-=\frac{2}{3}\frac{\varepsilon_1}{L^3}\bigg(\frac{\gamma_2^2}{\gamma_1^2}-1\bigg)z_2n_{2*}, %
%   S_0 = \ln\Gamma(z_1+1) -\ln\Gamma(N+1)-\ln\Gamma\big[z_1-N+1\big], %
}\vspace{3mm}\\ %
\displaystyle{\hspace{0mm}%
  \Delta S=S_+-S_-=\ln\frac{\Gamma(z_2+1)\,\Gamma(N+1)\,\Gamma(z_1-N+1)}
{\Gamma(z_2n_{2*}+1)\,\Gamma(N+1-z_2n_{2*})\,\Gamma(z_2+1-z_2n_{2*})\,\Gamma(z_1-N+1+z_2n_{2*})}, %
}\vspace{3mm}\\ %
\displaystyle{\hspace{0mm}%
  \Delta C_V=C_{V+}-C_{V-}=\frac{\tilde{Z}_0^2}{\tilde{Z}_1}, \quad %
  \Delta\beta_V=\beta_{V+}-\beta_{V-}=\frac{\tilde{Z}_0^2}{\tilde{Z}_1\varepsilon_1} %
  \bigg[N+\!\bigg(\frac{\gamma_2^2}{\gamma_1^2}-1\bigg)z_2n_{2*}\bigg]^{\!-1}. %
}%
\end{array}
\end{equation}
Due to the fact that at $\tau\rightarrow\tau_*+0$ the coefficient
$A_V\rightarrow -0$, the isobaric heat capacity tends to infinity
\begin{equation} \label{55}
\begin{array}{c}
\displaystyle{\hspace{0mm}%
  C_p\approx\frac{\tilde{Z}_0^2}{\tilde{Z}_1}+ %
  \big(\gamma_2^2-\gamma_1^2\big)\frac{\tilde{Z}_0^3}{\tilde{Z}_1^2\tilde{A}_1} %
  \frac{\tau_*}{\big(\tau-\tau_*\big)},
}%
\end{array}
\end{equation}
where
$\displaystyle{\tilde{A}_1}\equiv\frac{5}{2}\big(\gamma_1^2z_1n_{1*}+\gamma_2^2z_2n_{2*}\big)+ %
\big(\gamma_2^2-\gamma_1^2\big)\frac{\tilde{Z}_0}{\tilde{Z}_1} %
\bigg(\frac{5}{2}-\frac{\tilde{Z}_0\tilde{Z}_2}{\tilde{Z}_1^2}\bigg)$. %
The heat capacity goes to infinity due to the fact that the
interaction between particles is neglected. Taking into account such
interaction should lead to a finite value of the heat capacity at
$\tau_*$. This situation is qualitatively similar to that which
occurs in the boson gas near the condensation temperature
\cite{YP,YP2}. The coefficient of volumetric expansion $\alpha_p$
and the coefficient of isothermal compressibility $\gamma_T$ have
similar singularities:
\begin{equation} \label{56}
\begin{array}{c}
\displaystyle{\hspace{0mm}%
  \alpha_p\approx\frac{3}{2}\frac{\gamma_1^2}{\varepsilon_1} %
  \frac{\tilde{Z}_0^2}{\tilde{Z}_1\tilde{A}_1}\frac{\tau_*}{\big(\tau-\tau_*\big)}, \quad %
  \gamma_T\approx\frac{9}{4}\frac{\gamma_1^2}{\varepsilon_1} %
  \frac{L^3}{\tilde{A}_1}\frac{\tau_*}{\big(\tau-\tau_*\big)}. %
}%
\end{array}
\end{equation}

For an even smaller number of particles $N<N_m=0.061$ the
coefficient $A_V$ is always positive, so that the stability
condition $\big(\partial p/\partial V\big)_T<0$ for excited states
is not satisfied, and the system is in the ground state at any
temperature. A similar possibility was considered earlier by the
authors in the two-level system \cite{PS2}.

In the same way as is done for the case $0<N\le z_1$, the
temperature dependences of the populations and thermodynamic
characteristics can be analyzed for the case when at zero
temperature $M-1$ lower levels are completely filled and the level
$M$ can be filled partially. For definiteness, we briefly consider
the case when $z_1<N\le z_1+z_2$. Here at $T=0$ the first level is
completely filled with $n_1=1$, and on the second level there are
$N-z_1$ particles, so that $n_2=(N-z_1)/z_2$. The third and higher
levels are empty. In this case, there are two possibilities of
transition of the system to the excited state. At the temperature
\begin{equation} \label{57}
\begin{array}{c}
\displaystyle{\hspace{0mm}%
  \tau_{1\rightarrow 2}=\frac{\big(\gamma_2^2-\gamma_1^2\big)} %
  {\tilde{L}^2\Big[\theta\big((N-z_1)/z_2,z_2\big)-\theta(1,z_1)\Big]} %
}%
\end{array}
\end{equation}
the transition from the first level to the second level begins,
while the third level remains empty. A case is possible, when the
system is excited as a result of transition of particles from the
second to the third level with the ground level being completely
filled. This occurs at the temperature
\begin{equation} \label{58}
\begin{array}{c}
\displaystyle{\hspace{0mm}%
  \tau_{2\rightarrow 3}=\frac{\big(\gamma_3^2-\gamma_2^2\big)} %
  {\tilde{L}^2\Big[\theta(0,z_3)-\theta\big((N-z_1)/z_2,z_2\big)\Big]}. %
}%
\end{array}
\end{equation}
In fact, the excitation of the system will begin at that of the
temperatures (57),\,(58) which proves to be lower for a given number
of particles. The dependence of these temperatures on the number of
particles is shown in Fig.\,6.
\begin{figure}[h!]
\vspace{-2mm}  \hspace{0mm}
\includegraphics[width = 7.45cm]{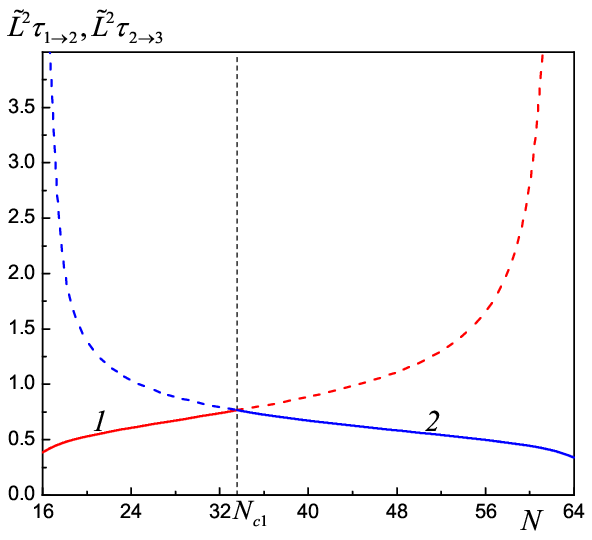} % 1.0\columnwidth
\vspace{-4mm} %
\caption{\label{fig06} %\hspace{10mm} %
Dependencies of characteristic temperatures (57),\,(58) on the
number of particles: %
$\tilde{L}^2\tau_{1\rightarrow 2}(N)$ -- {\it 1}, %
$\tilde{L}^2\tau_{2\rightarrow 3}(N)$ -- {\it 2}; $N_{c1}=33.55$. %
}\vspace{-1mm}%
\end{figure}

\section{Continual approximation} \vspace{-1mm}%%
If particles in a gas interact through short-range forces with
interaction radius $r_0$, then the ideal gas approximation can be
used if the volume per one particle $\upsilon=V/N$ is much larger
than the volume of an atom $\upsilon_0=4\pi r_0^3/3$, or
$\upsilon\gg \upsilon_0$. For a large system, the volume of which is
much larger than the volume per one particle $V\gg\upsilon$, the
condition $N\gg 1$ must be satisfied; however, it is necessary that
$N\ll\upsilon/\upsilon_0$. In this approximation, the extensive
quantities are proportional to the volume. In statistical mechanics,
it is common to go to the thermodynamic limit
$V/\upsilon_0\rightarrow\infty$ and $N\rightarrow\infty$ at
$\upsilon=V/N=\rm{const}$. In the problem under consideration, the
effects caused by the finite volume of the system are investigated.
Let us consider in more detail the transition to the continual
approximation at a large volume, when such effects are small.

Since in the number space $\big(n_x,n_y,n_z\big)$ there is a cube of
unit volume per one such state, then the total number of states in a
large system, for which the condition $n_x^2+n_y^2+n_z^2<\gamma^2$
is satisfied and whose energy is less than
$\varepsilon(\gamma)=\frac{\hbar^2}{2m}\left(\frac{2\pi}{L}\right)^2\gamma^2$,
is equal to the volume of the sphere
\begin{equation} \label{59}
\begin{array}{c}
\displaystyle{\hspace{0mm}%
  N(\gamma)=\frac{4\pi}{3}\gamma^3.
}%
\end{array}
\end{equation}
The number of states in the interval
$\,\gamma\div\gamma+\Delta\gamma\,$ is $\Delta
N(\gamma)=4\pi\gamma^2\Delta\gamma$, so that the density of the
number of states on the surface of a sphere of radius $\gamma$ is
equal to the surface area of this sphere
\begin{equation} \label{60}
\begin{array}{c}
\displaystyle{\hspace{0mm}%
  \upsilon(\gamma)=4\pi\gamma^2. %
}%
\end{array}
\end{equation}
The degeneracy factor of the level with energy
$\varepsilon(\gamma_j)$ with account of the two-fold degeneracy in
the spin projection is
\begin{equation} \label{61}
\begin{array}{c}
\displaystyle{\hspace{0mm}%
  z_j=2\upsilon(\gamma_j)=8\pi\gamma_j^2. %
}%
\end{array}
\end{equation}
Let us proceed to the description of the system in the continual
approximation. First assume that the number of levels $M$ is finite
and denote $k_j=(2\pi/L)\gamma_j$. Then the total number of particles %
\begin{equation} \label{62}
\begin{array}{c}
\displaystyle{\hspace{0mm}%
  N=\big(2L^2/\pi\big)\sum_{j=1}^Mk_j^2\,n(k_j). %
}%
\end{array}
\end{equation}
Further divide the interval of change of $k_j$ into equal intervals
$\Delta k\equiv (k_M-k_1)/(\gamma_M-\gamma_1)=2\pi/L$. Given the
definition of the de Broglie thermal wavelength
\begin{equation} \label{63}
\begin{array}{c}
\displaystyle{\hspace{0mm}%
  \Lambda\equiv\bigg(\frac{2\pi\hbar^2}{mT}\bigg)^{\!\!1/2}, %
}%
\end{array}
\end{equation}
let us introduce dimensionless quantities
$\Delta\kappa\equiv\Lambda\Delta k=2\pi(\Lambda/L)$, and
$\kappa_j=\Lambda k_j$. Then formula (62) will take the form
\begin{equation} \label{64}
\begin{array}{c}
\displaystyle{\hspace{0mm}%
  N=\big(L^3/\pi^2\Lambda^3\big)\sum_{j=1}^M\kappa_j^2\,n\bigg(\frac{\kappa_j}{\Lambda}\bigg)\Delta\kappa. %
}%
\end{array}
\end{equation}
Let us consider the limiting case $\Delta\kappa=2\pi(\Lambda/L)\ll 1$. %
In the limit $L\rightarrow\infty$ this condition is true at any
temperature. In the case of a finite volume system we are interested
in, this condition is satisfied if $\Lambda\ll L/2\pi$ or
$\sqrt{T}\gg\frac{2\pi\hbar}{L}\left(\frac{2\pi}{m}\right)^{\!1\!/2}$.
Thus, at a finite volume, one can move on to a continual description
at high temperatures when the de Broglie thermal wavelength is much
smaller than the cube's edge length. Considering $M\rightarrow\infty$ %
and passing from summation to integration in (64), we obtain
\begin{equation} \label{65}
\begin{array}{c}
\displaystyle{\hspace{0mm}%
  N=\big(L^3/\pi^2\Lambda^3\big)\int_0^\infty\!\kappa^2\,n\bigg(\frac{\kappa}{\Lambda},t\bigg)d\kappa= %
  \big(L^3/\pi^2\big)\int_0^\infty\!k^2n(k,t)dk.
}%
\end{array}
\end{equation}
Similarly, the transition to the continual approximation can be
performed for other thermodynamic quantities.

The equation that determines the average number of particles in each
state (13) can be written in the form
\begin{equation} \label{66}
\begin{array}{l}
\displaystyle{%
   \psi\big[z_j(1-n(k_j,t))+1\big] -\psi\big(z_jn(k_j,t)+1\big) = \frac{\big(\Lambda k_j\big)^2}{4\pi}-t,  %
}
\end{array}
\end{equation}
where $t\equiv\mu/T$. In the continual approximation $k_j\rightarrow
k$ can be considered as a continuous variable, and according to
formula (61)
\begin{equation} \label{67}
\begin{array}{l}
\displaystyle{%
  z_j\rightarrow z=\frac{2}{\pi}\big(Lk\big)^2. %
}%
\end{array}
\end{equation}
In this case, the quantities (35) take the form
\begin{equation} \label{68}
\begin{array}{cc}
\displaystyle{%
 g_2=\frac{L^3}{(2\pi)^3}\int_0^\infty\frac{k^2dk}{\theta^{(1)}(k,n)}, \quad %
 g_4=\frac{L^5}{(2\pi)^5}\int_0^\infty\frac{k^4dk}{\theta^{(1)}(k,n)}, \quad %
 \frac{1}{\theta^{(1)}}=\frac{L}{2\pi}\int_0^\infty\frac{dk}{\theta^{(1)}(k,n)}, %
}\vspace{3mm}\\ %
\displaystyle{\hspace{0mm}%
  d=\frac{L^5}{4\pi^4}\int_0^\infty n(k,t)k^4dk, %
}%
\end{array}
\end{equation}
where %\newpage
\begin{equation} \label{69}
\begin{array}{l}
\displaystyle{%
  \theta^{(1)}(k,n)=\psi^{(1)}\big[z(1-n)+1\big] +\psi^{(1)}\big(zn+1\big).  %
}
\end{array}
\end{equation}
If the conditions $z(1-n)\gg 1$ and $zn\gg 1$ are fulfilled, then
the distribution function turns into the usual Fermi-Dirac distribution %
\begin{equation} \label{70}
\begin{array}{l}
\displaystyle{%
  n_{F\!D}(k,t)=\frac{1}{e^{(k\Lambda)^2\!/4\pi\,-t}+1}. %
}%
\end{array}
\end{equation}
Figure 7 shows the distribution function in the continual
approximation $n(k,t)$ and the Fermi-Dirac function $n_{F\!D}(k,t)$.
In the main region of variation of the wave number $k$ these
functions are very close. The fundamental difference between them is
that $n_{F\!D}(k,t)$ is different from zero and unity in the entire
region $0\le k <\infty$, while $n(k,t)$ varies in the finite region
$k_1\le k\le k_2$, and furthermore at $k\le k_1$ it turns to unity
and at $k\ge k_2$ to zero.
\begin{figure}[h!]
\vspace{-1mm}  \hspace{0mm}
\includegraphics[width = 7.30cm]{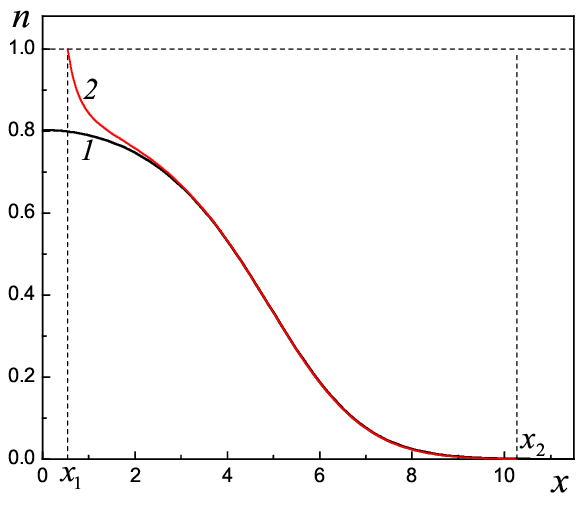} % 1.0\columnwidth
\vspace{-4mm} %
\caption{\label{fig07} %\hspace{10mm} %
The distribution functions: Fermi-Dirac $n_{F\!D}(x)$ (70) -- {\it 1}; %
in the continual approximation $n(x)$ by formula (66), %
for $t=1.4$, $L/\Lambda=3$ -- {\it 2}; %
$x\equiv k\Lambda$, $x_1=k_1\Lambda=0.55$, $x_2=k_2\Lambda=10.26$.
}\vspace{-1mm}%
\end{figure}

In the continual approximation, using the Fermi-Dirac distribution
(70), all thermodynamic quantities can be expressed through the
standard functions introduced by Stoner \cite{DS}:
\begin{equation} \label{71}
\begin{array}{l}
\displaystyle{%
  \Phi_s(t)=\frac{1}{\Gamma(s)}\int_0^\infty\!\frac{z^{s-1}dz}{e^{z-t}+1}, %
}%
\end{array}
\end{equation}
where $\Gamma(s)$ is the gamma function, $t\equiv\mu/T$. As a rule,
it is sufficient to use functions with indices $s=1/2,3/2,5/2$, for
which $\Gamma(1/2)=\sqrt{\pi},\,\Gamma(3/2)=\sqrt{\pi}/2,\,
\Gamma(5/2)=3\sqrt{\pi}/4$. A view of these functions is shown in
Fig.\,8. Let us give expressions through functions (71) for the
thermodynamic potential, number of particles, pressure, energy and
entropy of the Fermi gas in a large volume without taking into
account boundary effects:
\begin{equation} \label{72}
\begin{array}{cc}
\displaystyle{%
 \Omega_\infty=-\frac{2VT}{\Lambda^3}\,\Phi_{5/2}, \quad %
 N_\infty=\frac{2V}{\Lambda^3}\,\Phi_{3/2}, \quad %
 p_\infty=\frac{2T}{\Lambda^3}\,\Phi_{5/2},  %
}\vspace{3mm}\\ %
\displaystyle{\hspace{0mm}%
 E_\infty=\frac{3VT}{\Lambda^3}\,\Phi_{5/2}, \quad %
 S_\infty=\frac{2V}{\Lambda^3}\,\overline{\Phi}, %
}%
\end{array}
\end{equation}
where $\Phi_s\equiv\Phi_s(t)$,
$\overline{\Phi}\equiv\overline{\Phi}(t)\equiv\frac{5}{2}\,\Phi_{5/2}-t\,\Phi_{3/2}$.
Through the same functions one can express the heat capacities
\begin{equation} \label{73}
\begin{array}{cc}
\displaystyle{%
 C_{V\!\infty}=\frac{15}{2}\frac{V}{\Lambda^3}\bigg(\Phi_{5/2}-\frac{3}{5}\frac{\Phi_{3/2}^2}{\Phi_{1/2}}\bigg), \quad %
 C_{p\infty}=\frac{25}{2}\frac{V}{\Lambda^3}\Phi_{5/2}\bigg(\frac{\Phi_{1/2}\Phi_{5/2}}{\Phi_{3/2}^2}-\frac{3}{5}\bigg), %
}\vspace{3mm}\\ %
\displaystyle{\hspace{0mm}%
 C_{p\infty}}-C_{V\!\infty}=\frac{25}{2}\frac{V}{\Lambda^3}\frac{\Phi_{3/2}^2}{\Phi_{1/2}}\bigg(\frac{\Phi_{1/2}\Phi_{5/2}}{\Phi_{3/2}^2}-\frac{3}{5}\bigg)^{\!\!2}   %
\end{array}
\end{equation}
and the thermodynamic coefficients
\begin{equation} \label{74}
\begin{array}{cc}
\displaystyle{%
 \alpha_{p\infty}=\frac{5}{2T}\bigg(\frac{\Phi_{1/2}\Phi_{5/2}}{\Phi_{3/2}^2}-\frac{3}{5}\bigg), \quad %
 \gamma_{T\infty}=\frac{\Lambda^3}{2T}\frac{\Phi_{1/2}}{\Phi_{3/2}^2},\quad %
 \beta_{V\!\infty}=\frac{5}{2T}\frac{\Phi_{3/2}^2}{\Phi_{1/2}\Phi_{5/2}}\bigg(\frac{\Phi_{1/2}\Phi_{5/2}}{\Phi_{3/2}^2}-\frac{3}{5}\bigg). %
}%
\end{array}
\end{equation}

\begin{figure}[t!]
\vspace{-0mm}  \hspace{0mm}
\includegraphics[width = 7.30cm]{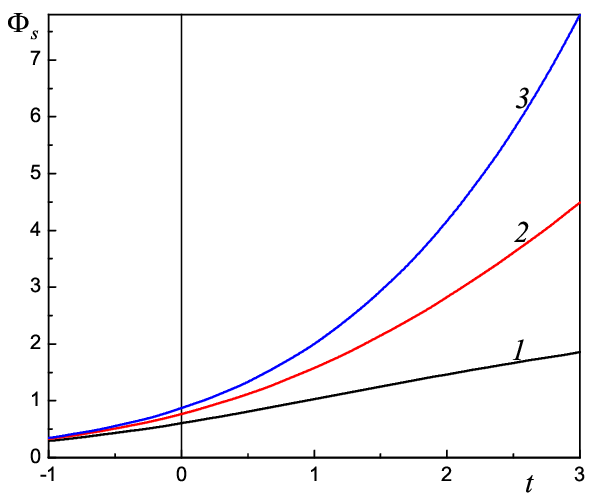} % 1.0\columnwidth
\vspace{-4mm} %
\caption{\label{fig08} %\hspace{10mm} %
The functions $\Phi_s(t)$ for the values: %
$s=1/2$ -- {\it 1}, $s=3/2$ -- {\it 2}, $s=5/2$ -- {\it 3}. %
}\vspace{-1mm}%
\end{figure}

Using asymptotic formulas, one can obtain thermodynamic relations in
the limiting cases of high and low temperatures. In the
high-temperature limit $t\rightarrow -\infty$ and $e^t\ll 1$, there
holds the asymptotics $\Phi_s(t)\approx e^{t}$. Another limiting
case $t\rightarrow +\infty$ corresponds to the case of a quantum gas
close to degeneracy at low temperatures. Here, with accuracy up to
exponentially small terms, we have the formulas
\begin{equation} \label{75}
\begin{array}{cc}
\displaystyle{%
 \Phi_{1/2}(t)=\frac{2t^{1/2}}{\sqrt{\pi}}\bigg(1-\frac{\pi^2}{24t^2}\bigg), \quad %
 \Phi_{3/2}(t)=\frac{4t^{3/2}}{3\sqrt{\pi}}\bigg(1+\frac{\pi^2}{8t^2}\bigg), \quad %
 \Phi_{5/2}(t)=\frac{8t^{5/2}}{15\sqrt{\pi}}\bigg(1+\frac{5\pi^2}{8t^2}\bigg). %
}%
\end{array}
\end{equation}
The given formulas are valid both in the limit of infinite volume
$V\rightarrow\infty$ and for a system of finite dimensions provided
the condition $\Lambda\ll L/2\pi$ is fulfilled.

\section{Conclusion} \vspace{-2mm}%%
In the presented work, we construct the thermodynamics and calculate
the energy, pressure, entropy, heat capacities and thermodynamic
coefficients of the Fermi gas enclosed in a cubic cavity of finite
volume. It is assumed that the average number of particles in a
volume can be arbitrary, in particular small and fractional. It is
taken into account that the form of the distribution function, as
previously shown by the authors \cite{PS,PS2}, differs from the
standard Fermi-Dirac function. The main feature of the exact
distribution function is that it does not have "exponential tails",
but turns to zero or unity at finite energy. The case of low
temperatures is considered, when the discrete structure of the
energy spectrum arising due to the size effect becomes important. It
is shown that the excitation of the system begins at some finite
temperature. Below this temperature the system continues to remain
in the same state as at zero temperature. It is characteristic that
if there is a not completely occupied level in the system, then the
entropy remains different from zero at zero temperature, so that the
third law of thermodynamics proves to be fulfilled only in the
Nernst formulation. As each new level begins filling, there happens
a jump in heat capacities. The transition to the continual
approximation is considered, which becomes possible if the de
Broglie thermal wavelength turns out to be much smaller than the
cube's edge length. In this limit, if the Fermi-Dirac function is
used as the distribution function, all thermodynamic quantities can
be expressed through the standard Fermi-Stoner functions.


\newpage
\begin{thebibliography}{99}
\bibitem{LL}
L.D.\,Landau, E.M.\,Lifshitz, {\it Statistical physics}: Vol.\,5
(Part\,1), Butterworth-Heinemann, 544\,p. (1980).
\bibitem{PS}
Yu.M.\,Poluektov and A.A.\,Soroka, {\it Quantum distribution
functions in systems with an arbitrary number of particles} (2023).
arXiv:2311.03003\,[quant-ph]
\bibitem{PS2}
Yu.M.\,Poluektov and A.A.\,Soroka, {\it On the thermodynamics of
two-level Fermi and Bose nanosystems}, Opt. Quant. Electron. \textbf{56}, 1349 (2024). %
doi:10.1007/s11082-024-07266-x; arXiv:2405.02427\,[quant-ph] %
\bibitem{Fr}
H.\,Fr\"{o}hlich, {\it The specific heat of electrons in small metal
particles at low temperatures}, Physica \textbf{4}, 406\,--\,412
(1937). doi:10.1016/S0031-8914(37)80143-3
\bibitem{PS3}
Yu.M.\,Poluektov and A.A.\,Soroka, {\it Thermodynamics of the Fermi
gas in a quantum well}, East Eur. J. Phys. \textbf{3}, \textnumero 4, 4\,--\,21 (2016). %
arXiv:1608.07205\,[cond-mat.stat-mech]
\bibitem{PS4}
Yu.M.\,Poluektov and A.A.\,Soroka, {\it Thermodynamics of the Fermi
gas in a nanotube}, East Eur. J. Phys. \textbf{4}, \textnumero 3, 4\,--\,17 (2017). %
arXiv:1704.03317\,[physics.gen-ph]
\bibitem{AS}
M.\,Abramowitz, I.\,Stegun (Editors), {\it Handbook of mathematical
functions, National Bureau of Standards Applied mathematics Series}
\textbf{55}, 1046\,p. (1964).
\bibitem{RR}
Yu.B.\,Rumer,  M.S.\,Ryvkin, {\it Thermodynamics, statistical
physics, and kinetics}, Mir publishers, Moscow, 600\,p. (1980). %
\bibitem{DS}
J.\,McDougall, E.C.\,Stoner, {\it The computation of Fermi-Dirac
functions}, Phil. Trans. Roy. Soc. A\,\textbf{237}, 67\,--\,104 (1938). %
doi:10.1098/rsta.1938.0004
\bibitem{YP}
Yu.M.\,Poluektov, {\it Isobaric heat capacity of an ideal Bose gas},
Russ. Phys. J. \textbf{44}\,(6), 627\,--\,630 (2001). %
doi:10.1023/A:1012599929812
\bibitem{YP2}
Yu.M.\,Poluektov, {\it A simple model of Bose-Einstein condensation
of interacting particles}, J. Low Temp. Phys. \textbf{186}, 347\,--\,362 (2017). %
doi:10.1007/s10909-016-1715-5
\end{thebibliography}
\end{document}